\documentclass[letterpaper,conference]{IEEEtran}
\usepackage[linesnumbered, ruled]{algorithm2e}
\usepackage[utf8]{inputenc}
\usepackage{graphicx}

\begin{document}

\title{An Adaptive Flow-Aware Packet Scheduling Algorithm for Multipath
  Tunnelling}

\author{\IEEEauthorblockN{Richard Sailer, Jörg Hähner}
\IEEEauthorblockA{Organic Computing Group\\
University of Augsburg, Augsburg, Germany\\
Email: richard.willi.sailer@student.uni-augsburg.de, joerg.haehner@informatik.uni-augsburg.de}
}

\maketitle

\begin{abstract}
This paper proposes AFMT, a packet scheduling algorithm to achieve adaptive
flow-aware multipath tunnelling. AFMT has two
unique properties. Firstly, it implements robust adaptive traffic splitting for
the subtunnels. Secondly, it detects and schedules bursts of packets cohesively,
a scheme that already enabled traffic splitting for load balancing with little
to no packet reordering.

Several NS-3 experiments over different network topologies show that AFMT
successfully deals with changing path characteristics due to background traffic while
increasing throughput and reliability.
\end{abstract}  

\section{Introduction}

Network paths can be unreliable and slow
{\cite{tanenbaum2003computer,Dominikn2011}}. Redundancy is an obvious solution
for this issue. Redundancy has been used to provide reliability and higher
throughput in many fields of computer engineering, e.g. databases, storage,
power supply, but seldomly for network paths. There are several proposed
concepts for redundancy: Multipath TCP {\cite{RFC6824}}
and Multipath Tunnelling {\cite{bednarek2016multipath}}.

\begin{figure}
  \center
  \includegraphics[width=0.8\linewidth]{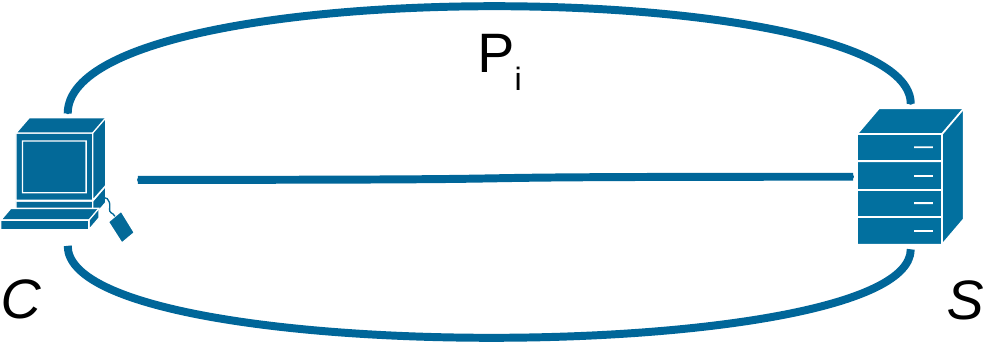}
  \caption{Multipath TCP Network Topology.
For every packet $p$ sent from client $C$ to server $S$, $C$ decides (schedules) the path $P_i$
to use. For this, $C$ and $S$ need an implementation of MPTCP and direct
access to all the paths.}
\label{fig:mptcp}

\end{figure}

Multipath TCP (MPTCP) grants more reliability, but needs every client and
every server to have direct full access to all network uplinks (figure
\ref{fig:mptcp}) {\cite{RFC6824}}. This complicates wiring. Additionally, it
needs all clients to implement MPTCP, a complex protocol. It also does not
solve the problem for UDP flows. So while MPTCP performs better than load
balancing, it still leaves a lot of issues unaddressed.

\begin{figure}
  \includegraphics[width=\linewidth]{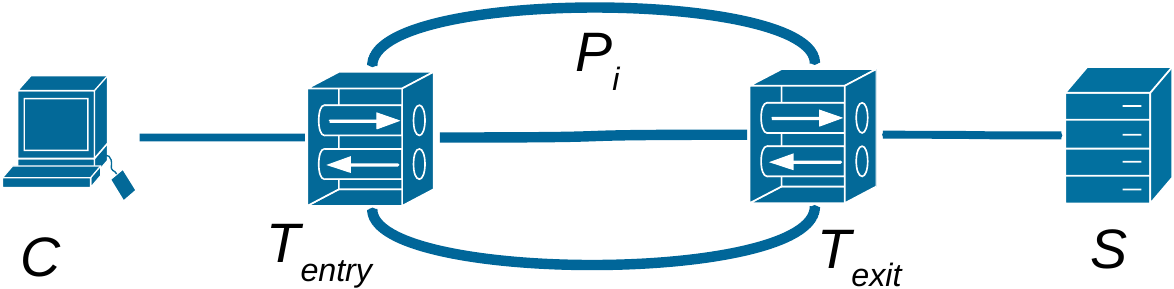}
  \caption{Multipath Tunnelling Network Topology.
A Packet $p$ sent from $C$ is encapsulated at the tunnel entry
$T_{entry}$ and sent via $P_i$ to the tunnel exit
$T_{exit}$. There, it is decapsulated and sent to $S$.}
  \label{fig:MT}
\end{figure}

Multipath Tunnelling (MT) addresses these issues. As visible in figure
\ref{fig:MT}, only the tunnel endpoints $T_{entry}$ and
$T_{exit}$ need to see and understand the sub-tunnels. The clients and
servers don't know they're connected by multiple paths, they need no
additional wiring or implementation of a new network protocol. Since all flows
between the two networks are tunnelled, this also works for UDP. Lastly, all
current prototypes and their concepts are less complex than the Multipath TCP
concept and its implementations {\cite{Paasch2014}}.

Our novel contributions include: We introduce AFMT, a packet
scheduling algorithm for adaptive flow-aware multipath tunnelling. AFMT
aims to overcome the throughput, reordering and adaptivity issues of existing MT
approaches. We evaluate AFMT for diverse networks with changing path
characteristics. This shows that AFMT improves reliability and throughput
compared to a classic packet scheduling scheme. The evaluation in this paper focuses
on wired connections, nevertheless AFMT research focuses on providing a general solution.

\section{Background}

\begin{figure}
  \center
  \includegraphics[width=0.8\linewidth]{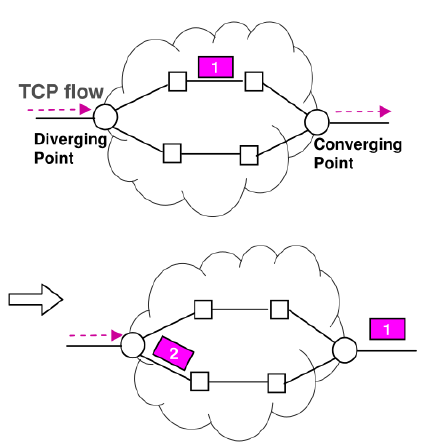}
  \caption{When an inter-flowlet delay $\delta$ is larger than the delay difference between
    two paths (MTBS), it is possible to send the two flowlets via the different
    paths and no packet reordering will occur {\cite{Kandula:2007:DLB:1232919.1232925}}.}
  \label{mtbs-diagram}
\end{figure}

TCP is sensitive to packet reordering {\cite{bohacek2006new}}, it interprets
reordering as a sign of packet loss and reacts with spurious retransmits and
throughput reduction. TCP uses a sliding window (congestion window,
{\textit{cwnd}}) algorithm to adjust its send rate to the path's capacity
{\cite{isi_793rfc81}}. Often, TCP sends the contents of a {\textit{cwnd}} as one
{\textit{burst}} or {\textit{flowlet}}
{\cite{Kandula:2007:DLB:1232919.1232925}}. The set of all transmitted packets in
a transport layer association is defined as a {\textit{flow}}
{\cite{tanenbaum2003computer}}. As illustrated in figure \ref{mtbs-diagram}, it
is possible to utilise flowlets to achieve traffic splitting with little to no
packet reordering.

Multipath tunnelling is known to induce heavy packet reordering
{\cite{bednarek2016multipath}}. This research proposes to reduce it
by using flowlet switching {\cite{Kandula:2007:DLB:1232919.1232925}}, a scheme
that reduced packet reordering for a similar problem (ISP load balancing) to
very little to no occurrence.

In multipath routing context, {\textit{packet scheduling}} refers to the task of
choosing an output queue for every packet from an input queue
{\cite{Paasch2014a}}. In a MT system one output queue maps to one
subtunnel. The behaviour of the packet scheduling algorithm is central to the
behaviour of the whole MT system {\cite{Dominikn2011}}.

\section{AFMT}

AFMT is built on the assumption that using these results gives comparable or
even better results than inventing an own path estimation scheme. DCCP and TCP
maintain an estimation of the path's RTT and capacity for their congestion
control. Decades of research have been invested to optimise these
\cite{tanenbaum2003computer,DCCP}. The following subsections describe the
realisation of flow awareness and adaptivity.

\subsection{Flow-Awareness}

%
%

To find the \textit{applicable subtunnels} and match the flow of a packet, a
central data structure that tracks all flows is necessary. Therefore AFMT uses a
\textit{flow table} with the \textit{flow\_id} as key and a tuple of last
subtunnel and timestamp as value. For every flow id it
returns the last subtunnel used by this flow and a timestamp. This timestamp
indicates the absolute time when the last packet of our flow was sent through
the associated subtunnel.

With this data AFMT can determine the \textit{applicable subtunnels} as shown in
Algorithm \ref{algo:flow-awareness}. Relevant entities are the subtunnels $s_i$,
the to-schedule packet $p$ and the absolute timepoints $t_i$. Initially AFMT
obtains the flow-id of $p$ ($p.flow\_id$). It can be obtained from the operating
system. Every operating system that supports network address translation (NAT)
needs to track flows and can (more or less directly) provide flow-ids. For Linux
this is possible relatively simple with the \textit{conntrack} netfilter module.
In most use cases AFMT targets, $T_{entry}$ is the internet gateway of a
organisation's local network, a device that already implements NAT. Therefore
there's no overhead for flow identification and tracking necessary.

Next we lookup the flow id in the flow table (Line 6). If it exists we assign
the data to two local variables (Line 7), and calculate $\delta$ the 
time that has passed since the last packet of our flow was sent.

Then the \textit{smoothed round trip time (SRTT)} of a subtunnel is obtained
from the transport protocol implementation\footnote{For Linux it is
  possible to get these values via getsockopt().}. It's more resistant to fluctuations
and therefore a more meaningful proposition about the path than the RTT\cite{isi_793rfc81}.

Knowing $s_i.SRTT$, the SRTT of a subtunnel $i$ it is possible to
predict when $p$ will arrive at $T_{exit}$, namely in $s_i.SRTT$ time from now
and $s_i.SRTT + \delta$ time from when $p_{last}$ was sent. Comparatively
$s_{last}.SRTT$ gives the arrival time of $p_{last}$ from when it was sent.
Therefore if $s_i.SRTT + \delta$ is larger than $s_{last}.SRTT$ $p$ will arrive
after $p_{last}$, and we can add $s_i$ to the list of
applicable subtunnels (Lines 11-12). This is done for all subtunnels other then
$s_{last}$.

After acquiring the list of applicable subtunnels AFMT selects the best of
them ($s_{opt}$) adaptivity wise, which is explained in more detail in the next
subsection (Line 15). Then the flow table is updated with the new values of
$s_{opt}$ and the current time $t_{now}$ and $p$ is finally sent via $s_{opt}$
(Lines 16-17).

If $p.flow\_id$ is not found in the flow table i.e. it starts a new flow, AFMT
directly calls the adaptive selection process with all available subtunnels
$s_1,...,s_n$ to determine $s_{opt}$ (Line 19). Then, as previously we update
the flow table and send $p$ via $s_{opt}$ (Line 20-21).

\begin{algorithm}
    \SetAlgoLined
    \DontPrintSemicolon
    Input\_Queue  $q$ \;
    Subtunnels $s_1,...,s_n$ \;
    \;
    $ p \leftarrow q.deque()  $ \;
    \;

    \eIf{$defined(flow\_table[p.flow\_id])$}{
      $(s_{last},\  t_{last}) \leftarrow flow\_table[p.flow\_id]$ \;
      $\delta = t_{now} - t_{last}$ \;

      $s_{applicable} \leftarrow $  $s_{last}$ \;

      \BlankLine

      \For {$s_i$ in other\_subtunnels} {
        \If {$s_i.SRTT + \delta > s_{last}.SRTT$} {
          $s_{applicable}$.append($s_i$)
        }
      }

      $s_{opt} \leftarrow select\_adaptively(s_{applicable})$ \;
       
      $flow\_table[p.flow\_id] \leftarrow \{s_{opt}, t_{now}\}$ \;
      $s_{opt}.send(p)$ \;
    }{ 
       
      $s_{opt} \leftarrow select\_adaptively(s_1,...,s_n)$ \;
      $flow\_table[p.flow\_id] \leftarrow \{s_{opt}, t_{now}\}$ \;
      $s_{opt}.send(p)$ \;
    }
    
\caption{AFMT: Flow-Awareness}
\label{algo:flow-awareness}    
\end{algorithm}

\subsection{Adaptivity}

Algorithm \ref{algo:adaptivity} illustrates how the best subtunnel adaptivity
wise is chosen. Line 1 iterates over all $s_i$ and calculates the
\textit{weighted fill} for each. Then, the subtunnel with the lowest one is
selected. \textit{weighted fill} aims to model the current load of the
subtunnel. It considers the fill of the buffer associated with $s_i$:
$s_i.fill$. $s_i.fill$ is added to the size of $p$ to get the full load this
subtunnel would have to shoulder. This value is divided by a value comparable to
the ``bandwidth-delay-quotient'': $s_i.cwnd / s_i.SRTT$.

\begin{algorithm}
    \SetAlgoLined
    \DontPrintSemicolon
      
      $s_{opt} \leftarrow s_{i}  $ with minimal $s_i.weighted\_fill$ \ \ \ \
      \ \ where \;
      $s_i.weighted\_fill = \frac{s_i.fill + p.size}{s_i.cwnd/s_i.SRTT}$ \;
 \Indp  \Indp \Indp \Indp \ \ \   $(= \frac{ s_i.fill + p.size  }{s_i.cwnd}* s_i.SRTT)$  \;

 \caption{AFMT: Adaptivity (\textit{select\_adaptively()})}
 \label{algo:adaptivity}
\end{algorithm}






\section{Evaluation}

To evaluate AFMT, it was implemented in NS-3 and used with several
network topologies and multiple payload flows.

\subsection{NS-3 and The Network}

For modelling wired links, NS-3 provides a PointToPoint and a CSMA model. 
We chose the CSMA model for all links since it provides the closest model
of the Ethernet and DSL links of a typical application case \cite{NS-3-model-library}. All CSMA net device
queues are configured as drop tail queues.

Data Rates of 16, 32 and 50 Mbit/s were chosen to model a fast cable internet
connection and two slower DSL subscriber lines. We also conducted a second set
of experiments with only two uplinks, modelling a 32 and a 50 Mbit/s line.
Intermediate Routers between $T_{entry}$ and $T_{exit}$ were introduced to
partially model the IP Layer routing overhead and different backbone paths of a
real client to online server path. All links in the backbone network are
modelled with 1 GBit/s. The same goes for the CSMA link from $T_{exit}$ to the
server, which represents a connection in a data center. The local network
configuration of the clients models a local gigabit Ethernet link. As a baseline
we also evaluated how the three payload flows perform if they are routed single
path via the fastest 50 Mbit/s link without any tunneling or multipath
aggregation.

For the CsmaChannel delay which models the propagation delay between two nodes
including all switches and hubs, we chose \textit{6560ns}. Considering an
average switch overhead of 600ns\cite{Barreiros2015} this models 1-2 switches
and 1-2 km of medium. Serialisation delay (sometimes called transmission delay)
and queuing delay are modelled by NS-3 based on the Channel Data Rates.
NS-3 does not simulate processing delays.

The simulation duration is 30 seconds. The payload flows start at second 4 and
cease at second 24. Between second 8 and 16 a background bulk TCP flow occurs
on the 32 Mbit/s uplink. This models a sudden decrease in the path's capacity to
observe how the different scheduling algorithms handle it.

To simulate application traffic we used the NS-3 packet-sink application on $C0-C2$
and three bulk send applications on $S$. This simulates three full speed
downloads via TCP. 
\subsection{AFMT and Round Robin Implementation}

For a first prototype we used TCP as transport protcol for the subtunnels.
Since the $T_{entry}$ and $T_{exit}$ nodes are under the control of the AFMT
system we can fully configure the TCP socket to benefit AFMT. The Delayed
acknowledgement extension allows TCP to only send an acknowledgement for every
second received data segment, or if a timeout occurs, to reduce overhead. NS-3
used a unusual high default of 200ms for the timeout, for accurate tunnel stats
we reduced it to the same value the Linux kernel uses: 40ms. TcpNoDelay was
enabled to get fast and interactive tunnel behaviour.

%
%
%
%
%
%
%

When tunnelling datagrams, TCP blurs the packet boundaries, since it's basically
a stream transport protocol\cite{tanenbaum2003computer}. To re-distinguish the
payload packets we introduced a small 8 byte header preceding every payload
datagram with its size. 

For comparison all experiments were also conducted with a round robin MT system
in the same network, with the same payload. The round robin (RR) scheduler used
UDP for it's subtunnels.





\subsection{General Results}

  \begin{table}
    \centering
    \begin{tabular}{|l|l|l|}
      \hline
      \textbf{Subtunnels} & \textbf{RR}  & \textbf{AFMT} \\ \hline \hline

      Three Subtunnels: 16, 32, 50 Mbit/s  & 63.02 & 105.89 \\ \hline
      
      Two Subtunnels: 32, 50 Mbit/s &  87.63 & 111 \\ \hline \hline
      
      No Tunnel, Single Path, 50 Mbit/s &  \multicolumn{2}{|c|}{52.7}  \\ \hline

    \end{tabular}
    \caption{Goodput over 30 seconds in MibiBytes for the different Scheduling Algorithms in
    different Network Topologies}
    \label{tab:Gooput-Results}
  \end{table}

The total goodput is shown in table \ref{tab:Gooput-Results}. All MT systems
provide a higher throughput than a single path solution with the fastest uplink.
The increase ranges from 21\% (three subtunnels RR) to 110\% (two subtunnels,
AFMT). For three subtunnels AFMT has a 60\% higher overall goodput with 105
Mibyte. However with two subtunnels the gap is much smaller with 27\%.

This indicates that the dynamic AFMT is better in dealing with diverse
paths and path capacity changes. The stable throughput additionally indicates
successful flow-aware traffic splitting. Low flow-awareness would have resulted
in packet reordering and therefore cwnd reduction and sudden throughput rate
drops. But both are considerabley reduced compared to RR as described in the
following two subsections.


\subsection{Three Subtunnels}

\begin{figure}
  \centering
  \includegraphics[width=\linewidth]{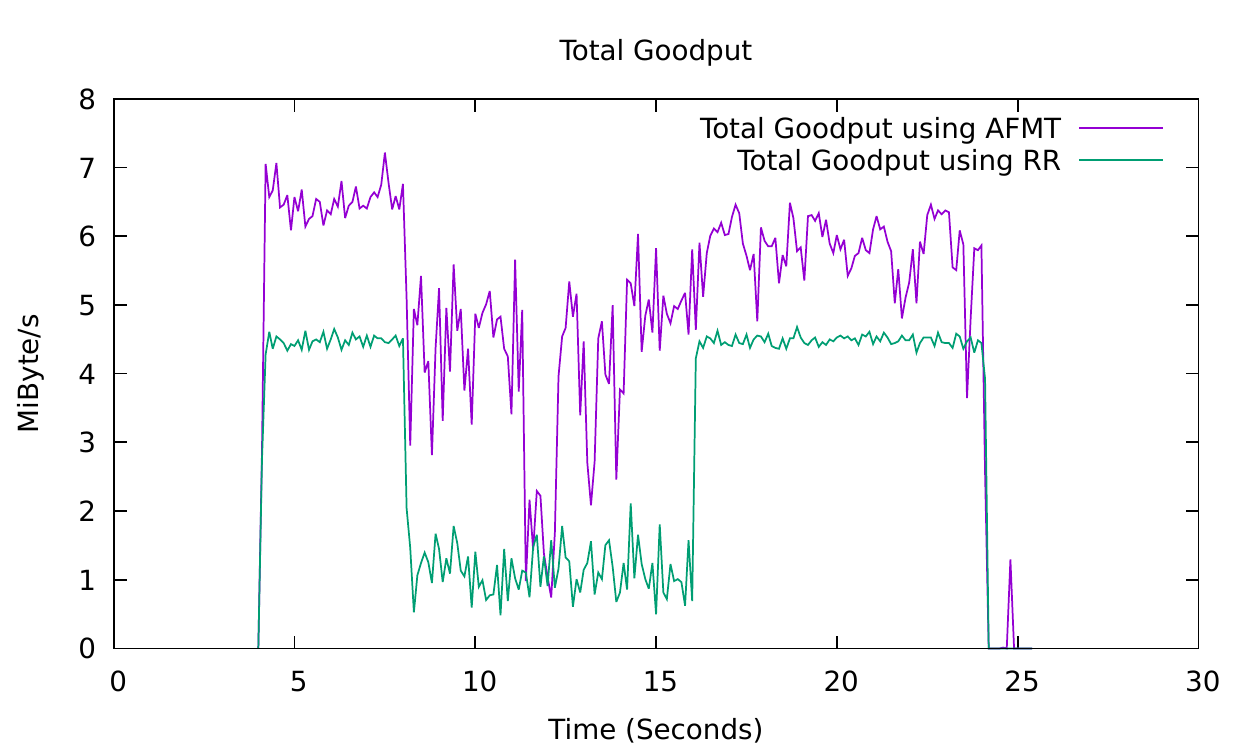}
  \caption{Total Goodput of AFMT System when using three subtunnels}
  \label{fig:total_goodput_afmt_32_50_16}
\end{figure}

For three subtunnels the overall goodput is plotted in Figure
\ref{fig:total_goodput_afmt_32_50_16}. AFMT has a
consistently higher throughput than RR. It starts at second
4 when the slow start algorithm of the subtunnels opens up the cwnd in the same
time the cwnds of the payload flows do. So no initial inertia is visible. After
that, until second 8 the goodputs lie around 6.5 MiB/s and 4.2 MiB/s.

At second 8 the gap widens as AFMT goodput drops to about 4 MiB/s and RR to
about 1. We assume this is because RR continues to send the same amount of
packets over the impaired path, which brings congestion and packet loss. At
second 12 AFMT drops down to the same goodput as RR. 1.5 seconds later it
recovers back to 4 MiB/s. At second 16 when the background traffic stops, both
systems recover. The AFMT goodput stays at an average of 5.8 MiB/s, while the RR
goodput stays at 4.1.

\subsection{Two Subtunnels}

\begin{figure}
  \centering
  \includegraphics[width=\linewidth]{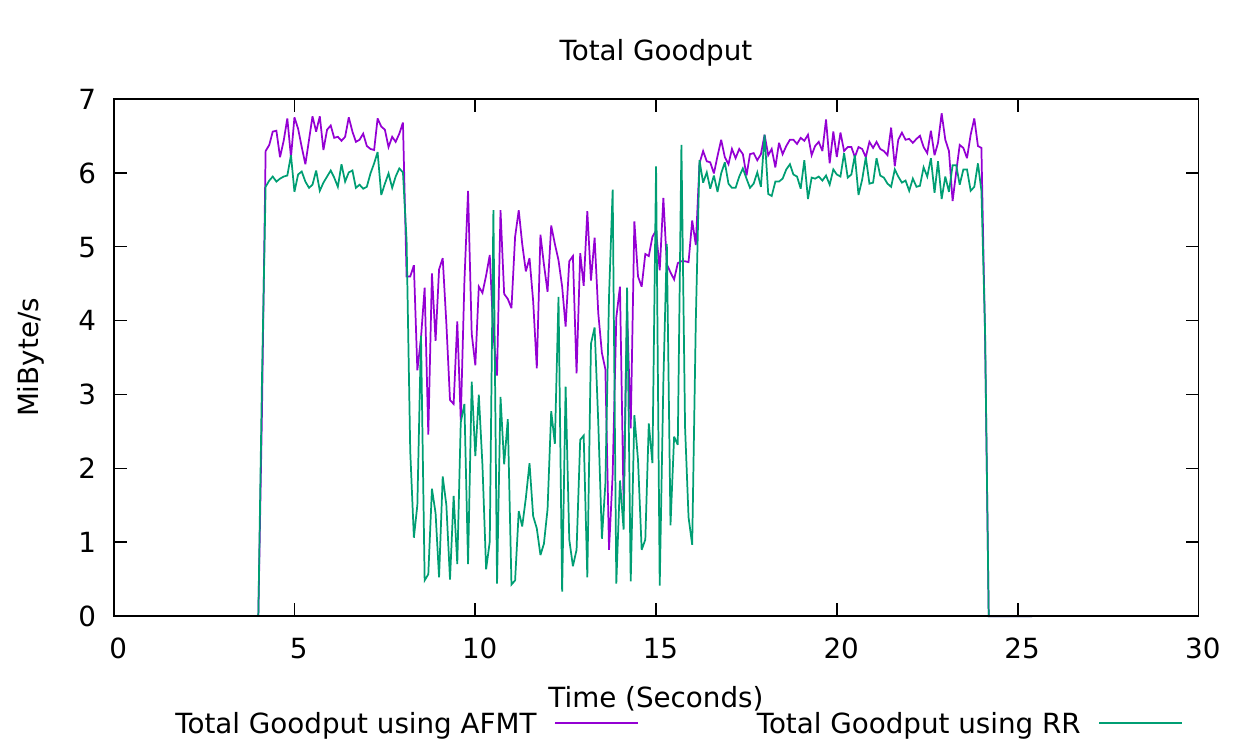}
  \caption{Total Goodput of AFMT System when using two subtunnels}
  \label{fig:total_goodput_afmt_32_50}
\end{figure}

The total goodput over time for both algorithms without the 16 Mbit/s subtunnel
is plotted in figure \ref{fig:total_goodput_afmt_32_50}. While AFMT opens its
cwnds still fast, the difference is smaller. AFMT transports ca. 6.5 MiB/s, 
RR oscillates around 5.8 MiB/s. At second 8 both throughputs drop and
again RR's throughput drops further to about 2 MiB/s compared to AFMT's 4 MiB/s.
However it is notable that for both systems the throughputs oscillate with
larger amplitudes than for three subtunnels. 
After second 16 both systems recover to their previous throughput rate. 

\section{Related Work}

\subsection{Multipath TCP}

A MPTCP scheduler has more options to avoid packet reordering than a MT
scheduler. It does not have to consider flowlets and
can define new flowlets suitable to avoid packet reordering. For adaptivity,
MPTCP trusts the congestion control of its subflows. Every time space in a
subflows' {\textit{cwnd}} opens and there is data to send, the scheduler is
invoked {\cite{Paasch2014}}.

LowRTT {\cite{Paasch2014a}} is a simple scheduler currently used as
default in the Linux Kernel. When invoked, it picks the subflow with the
lowest available RTT. It reduces head-of-line blocking and delay variation by about 20\%.

DPSAF {\cite{Xue2018}} is a sophisticated computation intensive scheduler for
vehicular networks. It tries to predict when the packets will arrive and sends
them out-of-order, so that they will arrive in order. While DPSAF might be a
good solution for vehicular networks with bad connectivity, it is unclear how
feasible it is for high speed internet usage.

\subsection{Multipath Tunnelling}

{\cite{bednarek2016multipath}} proposes a MT system for non-TCP loss-tolerant
media traffic and two subtunnel paths with fixed
characteristics. A DSL path with high stable bandwidth and a LTE path with low
varying bandwidth as overflow vault. It detects packet loss via sequence numbers
in a own header and adapts round robin weights acordingly. However, this reimplements
existing transport protocol functionality. It is not
flow-aware and has to re-reorders the packets with a reordering buffer.

{\cite{Dominikn2011}} researches multipath access networking in general.
Additionally the author designed a HTTP extension that splits videos in chunks
of fixed size and downloads them on separate paths. This needs changes of
the client application and only works for a specific case.

\section{Conclusions and Outlook}

In this paper we proposed AFMT a packet scheduling algorithm for multipath
tunnelling that increases throughput and reliability. Several NS-3 simulations
including changing path capacities have shown that AFMT effectively deals with
diverse and changing network paths.

These results were obtained although the experiments used the suboptimal TCP
protocol for the subtunnels. Our future work will evaluate and optimise AFMT
characteristics using DCCP with diverse payload traffic.

\bibliographystyle{IEEEtranS}
\bibliography{multipath_tunneling}

\end{document}